\documentclass[aps,prb,twocolumn,amssymb,showpacs,amsmath,superscriptaddress,floatfix]{revtex4}
\usepackage{graphicx}
\usepackage{bm}

\begin{document}
\title {Effects of the vortices and impurities on the nuclear spin relaxation rate in iron-based superconductors}

\author{Hong-Min Jiang}
\affiliation{Department of Physics, Hangzhou Normal University,
Hangzhou 310036, China} \affiliation{National Laboratory of Solid
State of Microstructure and Department of Physics, Nanjing
University, Nanjing 210093, China}
\author{Jia Guo}
\affiliation{National Laboratory of Solid State of Microstructure
and Department of Physics, Nanjing University, Nanjing 210093,
China}
\author{Jian-Xin Li}
\affiliation{National Laboratory of Solid State of Microstructure
and Department of Physics, Nanjing University, Nanjing 210093,
China}

\date{\today}

\begin{abstract}
The effects of magnetic vortices and nonmagnetic impurities on the
low energy quasiparticle excitations and the spin-lattice relaxation
rate are examined in the iron-based superconductors for the
$s_{\pm}$-, $s$- and $d$-wave pairing symmetries, respectively. The
main effect of the vortices is to enhance the quasiparticle
excitations and the spin-lattice relaxation rate for all symmetries,
and leads to a $T^{3}$ dependence of the relaxation rate followed by
a nearly $T$-linearity at lower temperatures. This enhancement can
only be seen for the $s_{\pm}$- and $d$-wave symmetries in the
presence of nonmagnetic impurities. These results suggest that the
$s_{\pm}$-wave and $d$-wave pairing states behave similarly in
response to the magnetic field and nonmagnetic impurities, therefore
it may be impossible to distinguish them on the basis of the
measurements of spin-lattice relaxation rates when a magnetic field
and/or impurity scatterings are present.
\end{abstract}

 \maketitle

\section{introduction}
Recently much attention has been payed to the newly discovered iron
arsenide superconductors,~\cite{kami1,xhchen1,zaren1,gfchen1,wang1}
which display superconducting transition temperature as high as more
than 50K, appear to share a number of general features with
high-$T_{c}$ cuprates, including the layered structure and proximity
to a magnetically ordered state.~\cite{kami1,cruz1,jdong} These
observations suggest that the conventional phonon-mediated pairing
mechanism appears to be unlikely and the magnetic correlations may
be relevant for superconductivity. So far, unconventional
superconductivity with pairing symmetry $s_{\pm}$ mediated by the
interband spin fluctuations has been proposed by a number of
theories for this layered iron
superconductors.~\cite{Mazin,kuroki,Yao,fczhang} Although, such a
popular proposal can explain some experimental findings, the
situation is complicated by the power-law temperature dependence of
the spin-lattice relaxation rate $T_{1}^{-1}\sim T^{n}$ below
$T_{c}$ with the doping dependent $n$ varying from 6 to
1.5.~\cite{hslee,grafe1,Hammerath1,Nakai1,Matano1,Kawasaki,Matano2,swzhang,Yashima,Fukazawa1,Fukazawa2}
Because a simple theoretical analysis shows that the relaxation rate
$T_{1}^{-1}$ should exhibit an exponential temperature dependence
for a fully gaped superconductors, while power law relation
$T_{1}^{-1}\sim T^{3}$ holds in the presence of line nodes in the SC
gap. Moreover, several experiments have reported the evidence of a
residual density-of-state (DOS) at zero energy in the SC state,
where the temperature dependent relaxation rate deviates from the
$T_{1}^{-1}\sim T^{3}$ relation and exhibits $T_{1}^{-1}\sim T$
behavior at very low temperature.~\cite{Nakai1,Michioka1,Hammerath1}
The disparity between the theoretical proposal and the experimental
fact presents a puzzle that should be resolved for the determination
of the SC pairing symmetry in these high-$T_{c}$ superconductors.

The relaxation rate has been studied in the presence of the
impurity-enhanced quasiparticle scattering at zero field in the band
representation, and it shows the power law temperature dependence
$T_{1}^{-1}\sim T^{3}$ for the
$s_{\pm}=\Delta_{0}\cos(k_{x})\cos(k_{y})$-wave SC
pairing.~\cite{Parker,YSenga,YBang} Although the enhancement of the
quasiparticle scattering has been well included in the previous
studies, the relationship between the SC gap and the impurities
addressed in the band representation were debated
controversially.~\cite{Onari1,Onari2} More importantly, actual
spin-lattice relaxation rate measured in NMR experiments are
conducted under a magnetic field of several Tesla, where the
relaxation rate shows the temperature dependence $T_{1}^{-1}\sim
T^{3}$ and even $T_{1}^{-1}\sim
T$.~\cite{Nakai1,Fukazawa1,lma,Michioka1} At the same time, H. J.
Grafe \textit{et al.} has found that the superconducting vortices
contribute to the spin-lattice relaxation rate when the magnetic
field is perpendicular to the conducting plane but not for the
parallel direction.~\cite{grafe1,Kitagawa1} In fact, it is unclear
up to now whether the different value of $n$ in the temperature
dependence of $T_{1}^{-1}$ is caused by a change of the pairing
symmetry with doping, a disorder scattering effect in an $s_{\pm}$
gap superconductor, or other unknown mechanism. In view of these
theoretical and experimental facts, it is necessary to compare and
contrast different pairing symmetries by studying the effects of
impurities and magnetic field on the spin-lattice relaxation rate in
the analysis of the standardized procedure extracting the gap
symmetry.

The purpose of this study is to present such a contrastive study on
three different superconducting (SC) pairing symmetries with
$s_{\pm}$-, $s$- and $d$-wave, respectively. For this purpose, we
calculate the DOS and the spin-lattice relaxation rate in the
Fe-based superconductors in the presence of magnetic field and
nonmagnetic impurities by self-consistently solving the
Bogoliubov-de Gennes equations based on the simple two-orbital
model. It is shown that the characteristic low energy quasiparticle
excitations depend on the gap functions in the presence of magnetic
vortices and nonmagnetic impurities. The magnetic vortices
contribute significantly to the spin-lattice relaxation rate and
lead to the relation $T_{1}^{-1}\sim T^{3}$ followed by a nearly
linear dependence at lower temperatures for all symmetries. While in
the presence of nonmagnetic impurities, this enhancement of
$T_{1}^{-1}$ can only be seen for the $s_{\pm}$- and $d$-wave
symmetries. Thus, in the presence of magnetic field and nonmagnetic
impurities, the $s_{\pm}$-wave pairing behaves much like what the
$d$-wave does.

The remainder of the paper is organized as follows. In Sec. II, we
introduce the model Hamiltonian and carry out analytical
calculations. In Sec. III, we present numerical calculations and
discuss the results. In Sec. IV, we make a conclusion.

\section{THEORY AND METHOD}
We start with an effective two-orbital model that takes only the
iron $d_{xz}$ and $d_{yz}$ orbitals into account.~\cite{ragh1} By
assuming an effective attraction that causes the superconducting
pairing and including the possible interactions between the two
orbitals' electrons, one can construct an effective model to study
the vortex and impurity physics of the iron-based superconductors in
the superconducting state:
\begin{eqnarray}
H=H_{0}+H_{pair}.
\end{eqnarray}
The first term is a tight-binding model
\begin{eqnarray}
H_{0}=&&-\sum_{ij,\alpha\beta,\sigma}e^{i\varphi_{ij}}t_{ij,\alpha\beta}
c^{\dag}_{i,\alpha,\sigma}c_{j,\beta,\sigma} \nonumber
\\ &&-\mu\sum_{i,\alpha,\sigma}c^{\dag}_{i,\alpha,\sigma}c_{i,\alpha,\sigma}
+\sum_{i,\alpha,\sigma}U_{i}c^{\dag}_{i,\alpha,\sigma}c_{i,\alpha,\sigma},
\end{eqnarray}
which describes the electron effective hoppings between sites $i$
and $j$ of the Fe ions on the square lattice, including the intra-
($t_{ij,\alpha\alpha}$) and inter-orbital ($t_{ij,\alpha,\beta},
\alpha\neq\beta$) hoppings with the subscripts $\alpha$, $\beta$
($\alpha,(\beta)=1,2$ for $xz$ and $yz$ orbital, respectively)
denoting the orbitals and $\sigma$ the spin.
$c^{\dag}_{i,\alpha\sigma}$ creates an $\alpha$ orbital electron
with spin $\sigma$ at the site $i$ ($i\equiv(i_{x},i_{y})$), and
$\mu$ is the chemical potential. The magnetic field is introduced
through the Peierls phase factor $e^{i\varphi_{ij}}$ with
$\varphi_{ij}=\frac{\pi}{\Phi_{0}}\int^{r_{i}}_{r_{j}}\mathbf{A(r)}\cdot
d\mathbf{r}$ in the vortex state, where $A=(-Hy, 0, 0)$ stands for
the vector potential in the Landau gauge and $\Phi_{0}=hc/2e$ is the
superconducting flux quantum. In the case of the SC state with
impurities, we randomly select the half number of total sites, at
where the random disorder potentials $U_{i}$ uniformly distributed
over [$-U,U$] are set in. Then $U$ is a parameter to characterize
the strength of the disorder. The hopping integrals are chosen as to
capture the essence of the density function theory (DFT)
results.~\cite{gxu1} Taking the hopping integral between the
$d_{yz}$ orbitals $|t_{1}|=1$ as the energy unit, we have,
\begin{eqnarray}
t_{i,i\pm \hat{x},xz,xz}=&&t_{i,i\pm \hat{y},yz,yz}=t_{1}=-1.0 \nonumber\\
 t_{i,i\pm\hat{y},xz,xz}=&&t_{i,i\pm \hat{x},yz,yz}=t_{2}=1.25 \nonumber\\
 t_{i,i\pm\hat{x}\pm\hat{y},xz,xz}=&&t_{i,i\pm\hat{x}\pm
 \hat{y},yz,yz}=t_{3}=-0.9 \nonumber\\
t_{i,i+\hat{x}-\hat{y},xz,yz}=&&t_{i,i+\hat{x}-\hat{y},yz,xz}
 =t_{i,i-\hat{x}+\hat{y},xz,yz} \nonumber\\
=&&t_{i,i+\hat{x}-\hat{y},yz,xz}=t_{4}=-0.85 \nonumber\\
t_{i,i+\hat{x}+\hat{y},xz,yz}=&&t_{i,i+\hat{x}+\hat{y},yz,xz}
 =t_{i,i-\hat{x}-\hat{y},xz,yz} \nonumber\\
=&&t_{i,i-\hat{x}-\hat{y},yz,xz}=-t_{4}.
\end{eqnarray}
Here, $\hat{x}$ and $\hat{y}$ denote the unit vector along the $x$
and $y$ direction, respectively.

The second term accounts for the superconducting pairing.
Considering that a main purpose here is to address the effects of
the magnetic vortices and nonmagnetic impurities on the spin-lattice
relaxation rate in the iron-based superconductors, we take a
phenomenological form for the intra-orbital pairing interaction,
\begin{eqnarray}
H_{pair}=&&\sum_{i
j,\alpha}V_{ij}(\Delta_{ij,\alpha\alpha}c^{\dag}_{i,\alpha\uparrow}c^{\dag}_{j,\alpha\downarrow}
+h.c.)
\end{eqnarray}
with $V_{ij}$ as the strengths of effective attractions.

Thus, we obtain the Bogoliubov-de Gennes equations for this model
Hamiltonian~\cite{hmjiang}
\begin{eqnarray}
\sum_{j,\alpha<\beta}\left(
\begin{array}{cccc}
H_{ij,\alpha\alpha,\sigma} &
\Delta_{ij,\alpha\alpha} & H_{ij,\alpha\beta,\sigma} & 0 \\
\Delta^{\ast}_{ij,\alpha\alpha} &
-H^{\ast}_{ij,\alpha\alpha,\bar{\sigma}} &
0 & -H^{\ast}_{ij,\alpha\beta,\bar{\sigma}} \\
H_{ij,\alpha\beta,\sigma} & 0 &
H_{ij,\beta\beta,\sigma} & \Delta_{ij,\beta\beta} \\
0 & -H^{\ast}_{ij,\alpha\beta,\bar{\sigma}} &
\Delta^{\ast}_{ij,\beta\beta} &
-H^{\ast}_{ij,\beta\beta,\bar{\sigma}}
\end{array}
\right)\nonumber\\
\times\left(
\begin{array}{cccc}
u^{n}_{j,\alpha,\sigma} \\
v^{n}_{j,\alpha,\bar{\sigma}} \\
u^{n}_{j,\beta,\sigma} \\
v^{n}_{j,\beta,\bar{\sigma}}
\end{array}
\right)= E_{n}\left(
\begin{array}{cccc}
u^{n}_{i,\alpha,\sigma} \\
v^{n}_{i,\alpha,\bar{\sigma}} \\
u^{n}_{i,\beta,\sigma} \\
v^{n}_{i,\beta,\bar{\sigma}}
\end{array}
\right),
\end{eqnarray}
where,
\begin{eqnarray}
H_{ij,\alpha\alpha,\sigma}=&&-e^{i\varphi_{ij}}t_{ij,\alpha\alpha}-\mu \nonumber\\
H_{ij,\alpha\beta(\beta\neq\alpha),\sigma}=&&-e^{i\varphi_{ij}}t_{ij,\alpha\beta(\beta\neq\alpha)}.
\end{eqnarray}
$u^{n}_{j,\alpha,\sigma}$ ($u^{n}_{j,\beta,\bar{\sigma}}$),
$v^{n}_{j,\alpha,\sigma}$ ($v^{n}_{j,\beta,\bar{\sigma}}$) are the
Bogoliubov quasiparticle amplitudes on the $j$-th site with
corresponding eigenvalues $E_{n}$.

The pairing amplitude and electron densities are obtained through
the following self-consistent equations,
\begin{eqnarray}
\Delta_{ij,\alpha\alpha}=&&\frac{V_{ij}}{4}\sum_{n}(u^{n}_{i,\alpha,\sigma}
v^{n\ast}_{j,\alpha,\bar{\sigma}}
+v^{n\ast}_{i,\alpha,\bar{\sigma}}u^{n}_{j,\alpha,\sigma})\times \nonumber\\
&&\tanh(\frac{E_{n}}{2k_{B}T})
\nonumber\\
n_{i,\alpha,\uparrow}=&&\sum_{n}|u^{n}_{i,\alpha,\uparrow}|^{2}f(E_{n}) \nonumber\\
n_{i,\alpha,\downarrow}=&&\sum_{n}|v^{n}_{i,\alpha,\downarrow}|^{2}[1-f(E_{n})].
\end{eqnarray}

The site-averaged DOS $N(E)$ is calculated by
\begin{eqnarray}
N(E)=&&
-\frac{1}{N}\sum_{i}\sum_{n,\alpha}[|u_{i,\alpha,\uparrow}^{n}|^{2}f^{'}(E_{n}-E) \nonumber\\
&&+|v_{i,\alpha,\downarrow}^{n}|^{2} f^{'}(E_{n}+E)],
\end{eqnarray}
where, $f^{'}(E)$ is the derivative of the Fermi-Dirac distribution
function with respect to energy. The nuclear spin-lattice relaxation
rate is given by~\cite{Takigawa}
\begin{eqnarray}
R(r_{i},r_{i'})=&&\textmd{Im}\chi_{+,-}(r_{i},r_{i'},i\Omega_{n}\rightarrow\Omega+i\eta)/(\Omega/T)|_{\Omega\rightarrow
0}\nonumber\\
=&&-\sum_{n,n'}[\mathcal {U}^{n}_{i}\mathcal
{U}^{n\ast}_{i'}\mathcal {V}^{n'}_{i}\mathcal {V}^{n'\ast}_{i'}
-\mathcal {V}^{n}_{i}\mathcal {U}^{n\ast}_{i'}\mathcal {U}^{n'}_{i}\mathcal {V}^{n'\ast}_{i'}]\nonumber\\
&&\times\pi T f'(E_{n})\delta(E_{n}-E_{n'}).
\end{eqnarray}
Here, $\mathcal {U}^{n}_{i}=u^{n}_{i,\alpha}+u^{n}_{i,\beta}$ and
$\mathcal {V}^{n}_{i}=v^{n}_{i,\alpha}+v^{n}_{i,\beta}$.  We choose
$\textbf{r}_{i}=\textbf{r}_{i'}$ because the nuclear spin-lattice
relaxation at a local site is dominant. Then the site-dependent
relaxation time is given by $T_{1}(r)=1/R(r,r)$ and the bulk
relaxation time $T_{1}=(1/N)\sum_{r}T_{1}(r)$.

In numerical calculations, $V_{ij}$ is chosen to give a short
coherence length of a few lattice spacing in the SC state being
consistent with experiments.~\cite{takeshita} Under the conditions
$V_{ij}\sim2.0$, $\mu=1.2$ at temperature $T=1\times10^{-5}$, the
filling factor
$n=\sum_{i,\alpha,\sigma}(n_{\alpha,\sigma})/(N_{x}N_{y})=1.9$ and
the coherent peak of the SC order parameter in the DOS is at
$\Delta_{max}\sim0.25$. Thus, we estimate the coherence length
$\xi_{0}\sim E_{F}a/|\Delta_{max}|\sim5a$,~\cite{ydzhu} with $a$
being the Fe-Fe distance on the square lattice. Due to this short
coherence length, presumably the system will be a type-II
superconductor. To study the vortex states, we employ the magnetic
unit cell with size $N_{x}\times N_{y}=48\times24$ that accommodates
two magnetic vortices, unless otherwise specified. In view of these
parameters, we estimate the upper critical field $B_{c2}\sim100T$.
Therefore, the model calculation is particularly suitable for the
iron-based type-II superconductors such as CaFe$_{1-x}$Co$_{x}$AsF,
Eu$_{0.7}$Na$_{0.3}$Fe$_{2}$As$_{2}$ and FeTe$_{1-x}$S$_{x}$, where
the typical coherence length $\xi_{0}$ deduced from the experiments
is of a few lattice spacing and the upper critical field achieves as
high as dozens of Tesla.~\cite{takeshita}

\section{results and discussion}

Since no final consensus on the SC pairing symmetry has yet
been achieved, we choose three possible singlet pairing symmetries,
i.e., the most popular sign-reversed $s_{\pm}$-wave, the on-site
$s$-wave and the $d$-wave symmetries with their respective gap
functions $\Delta_{s_{\pm}}=\Delta_{0}\cos(k_{x})\cos(k_{y})$,
$\Delta_{s}=\Delta_{0}$ and
$\Delta_{d}=\Delta_{0}[\cos(k_{x})-\cos(k_{y})]$ to carry out the
contrasting study. At the end of this section, we will also touch on
another possibility of the $s_{\pm}$-wave with angular variation
along the electron Fermi pockets, which will be referred to as
$s_{h}$-wave. In order to obtain comparable values of critical
temperature $T_{c}$ in the self-consistent calculations, we set
respectively $V_{ij}=1.6$ for $s_{\pm}$-wave, $V_{ij}=2.0$ for
$s$-wave, and $V_{ij}=1.8$ for $d$-wave pairing symmetries.

To begin with, we briefly summarize the site-averaged DOS spectra
$N(E)$ in the uniform SC state at $T=1\times10^{-5}$ as shown by the
solid lines in Fig. 1. For the $s_{\pm}$- and $s$-wave symmetries,
no node exists in the gap along the Fermi surfaces. Correspondingly,
the full gap structures can be seen in the DOS as shown in Figs.
1(a) and 1(b). In the $d$-wave symmetry, the gap function has line
nodes at $k_x=k_y$, which cross the hole pocket but do not intersect
the electron pocket. As a result, the DOS consists of a small
V-shaped gap structure at very low energy and a U-shaped gap
structure at higher energy as shown in Fig. 1(c), which exhibits a
difference from that in high-$T_c$ cuprates.

Next, we show the DOS in the magnetic vortex state and in the
presence of nonmagnetic impurities as presented by the dotted and
dashed lines in Fig. 1, respectively. In the vortex state, the
application of a magnetic field will induce the quasiparticle flow
around the vortex core, such that the nodal line will appear in the
otherwise fully gaped SC state due to an additional Doppler shift in
the quasiparticle energy, giving rise to the V-shaped gap structures
for the $s_{\pm}$- and $s$-wave symmetries, as shown in Figs. 1(a)
and 1(b). In contrast, in the presence of impurities, the DOS shows
the V-shaped structure for the $s_{\pm}$-wave symmetry but the
U-shaped for the $s$-wave symmetry [Figs. 1(a) and 1(b)]. This is
due to the fact that the impurity potential has intra- and
inter-band components for the multiband materials. The intra-band
components scatter fermions that have the same sign for the
$s_{\pm}$-wave SC order parameter and therefore do not affect the
superconductivity. Whereas the inter-band components scatter
fermions with opposite SC order parameters, thus have the pairing
breaking effect. As a result, they yield an obvious decrease in
$T_{c}$ and simultaneously introduce the V-shaped feature in DOS. In
the case of the $s$-wave symmetry, no sign change in the SC order
parameters occurs on both the electron and hole pockets, so that no
obvious pairing breaking effect is induced by the impurities and the
U-shaped DOS is untouched.

For the $d$-wave symmetry, the small V-shaped superconducting gap is
filled by the low energy quasiparticles induced by either vortices
or impurities, resulting in a pseudogap-like U-shaped feature with
finite density at Fermi energy, as indicated by the dotted and
dashed lines in Fig. 1(c). The line nodes existing on the
hole-pocket in the $d$-wave symmetry make it vulnerable to the
impurities, resulting in the disappearance of the small V-shaped
structure. However, the full gap opening on the electron-pocket is
robust against impurities, so that the pseudogap-like U-shaped
feature is obtained. In the vortex state, since the quasiparticles
induced by the vortex come preferably from the line nodes, the
pseudogap-like U-shaped structure remains. The results in Fig. 1
indicate that the characteristic low energy quasiparticle
excitations depend on the gap functions in the presence of magnetic
vortices and nonmagnetic impurities.

Now, we turn to the discussion of the temperature $(T)$ dependence
of the nuclear relaxation rate. For the uniform SC state, both the
$s_{\pm}$- and $s$-wave symmetries produce a power law relation
$T_{1}^{-1}\sim T^{5}$ below $T_{c}$ and it evolves into an
exponential dependence at very low temperature, as shown by the
solid lines in Figs. 2 and 3,~\cite{Vorontsov1} which are the
consequence of the full-gap DOS in Fig. 1. In the vortex state, due
to the similar V-shaped DOS for both the $s_{\pm}$- and $s$-wave
symmetries, $T_{1}^{-1}$ changes it's $T$ dependence to $T^{3}$
below $T_{c}$ and becomes nearly proportional to $T$ for the
$s_{\pm}$-wave symmetry while $T^{1.5}$ for the $s$-wave symmetry at
low temperature, as denoted by the dotted lines in Figs. 2(a) and
3(a), respectively. [The dotted lines in Figs. 2(a)-5(a) show the
results for magnetic unit cell with size $48\times24$, while the
dash-dotted lines in these figures the results for magnetic unit
cell with size $40\times20$.]

We note that both the $T^{3}$ dependence and the low-$T$ feature
with nearly linear slop are reminiscent of the experimental
observations,~\cite{Hammerath1,Nakai1} where the $T$-linear
dependence has been regarded as the evidence for a residual density
of states at zero energy in SC state. As mentioned above, although
the $s_{\pm}$-wave pairing is basically the full gap, the
application of a magnetic field will cause the quasiparticle flow
around the vortex core, such that the nodal line will appear in the
SC state due to an additional Doppler shift quasiparticle energy,
giving rise to the V-shaped DOS and $T^{3}$ dependent relaxation
rate. In view of this, we may expect that the slop of temperature
dependent $T_{1}^{-1}$ should be insensitive to the strength of the
magnetic field, which has already been observed in
experiment.~\cite{lma} This is evident by the comparison between the
dotted- and dash-dotted-curves shown in Fig. 2(a), where the
magnetic field for the dash-dotted-curve is about $1.5$ times as
large as that for the dotted-curve.

A striking difference between the $s_{\pm}$-wave and the $s$-wave
states in the $T$-dependence of $T_{1}^{-1}$ appears when the effect
of the impurity scattering is considered, as shown in Figs. 2(b) and
3(b). For the $s_{\pm}$-wave state, $T_{1}^{-1}$ deviates gradually
from $T^{5}$ to a overall $T^{3}$ behavior at weak disorder such as
$U=1$, then to a $T$ linear dependence at low temperatures as the
disorder strength $U$ is increased to about $U=1.5$, which depicts
the sensitivity of the SC order to the impurities and is in
accordance with the DOS results. On the other hand, $T_{1}^{-1}$
changes little upon the introduction of the disorder for the
$s$-wave symmetry. We also notice that $T_{c}$ is reduced
substantially for the $s_{\pm}$-wave symmetry at the moderate and
even the weak disorder strength, as can be seen from the lower-$T$
shift of the inflexion point on the curves in Fig. 2(b). This again
reflects the fact that the sign-reversed $s_{\pm}$-wave pairing is
fragile against the nonmagnetic impurities, which has also been
predicted theoretically by adopting a more sophisticated orbital
model and conformed experimentally in the specific heat and
resistivity measurements.~\cite{Onari1,Mu1,Guo1}

In Fig. 4, we present the results for the $d$-wave symmetry. Unlike
the formers, the overall $T$-dependence of $T_{1}^{-1}$ in the
uniform SC state roughly follows $T^{3}$ relation due to the
existing of nodal line in the gap structure. When a magnetic field
is applied, $T_{1}^{-1}$ exhibits a $T$-linear dependence at low temperatures,
though a $T^{3}$ behavior following $T_c$
still remains. We notice that this trend is rather robust against
the magnitude of the magnetic field, as the results for two cases
are nearly the same though their magnitude differs about $1.5$
times, as shown in Fig. 4(a). Thus, for the three different pairing
states, $T_{1}^{-1}$ exhibits nearly similar $T$-dependence in the
magnetic vortex state, due to the presence of quasiparticles induced
by the vortex. In the presence of impurities, $T_{1}^{-1}$ for both
the $d$-wave and the $s_{\pm}$-wave symmetries exhibit
a consecutive change from the $T^{3}$ to $T$-linear dependence when temperature is
decreased, which contrasts with that for the $s$-wave state.

We notice that the actual multiple Fermi surface sheets in
iron-based superconductors are not exactly reproduced by the
two-orbital model, so the general structure of the gap in the
$s_{\pm}$-wave channel may involve the angular variation along the
electron Fermi pockets and have the form
$\Delta_{h}=\Delta_{0}\{\cos(k_{x})\cos(k_{y})+h[\cos(k_{x})+\cos(k_{y})]\}$,
where the factor $h$ measures the strength of the angular dependent
variations along the electron Fermi
pockets.~\cite{Chubukov1,SMaiti1,SMaiti2,Basov1,AFKemper1} In the
case of $h>1$, there will be accidental nodes along the electron
Fermi pockets,~\cite{SMaiti3} which will be focused here. Such
accidental nodes are reflected in the V-shaped DOS in Fig. 1(d) and
nearly $T^{4}$ dependence of $T_{1}^{-1}$ for the typical results
with $h=1.5$ in the uniform SC state [Fig. 5]. This result differs
from the $T^{3}$ dependence for the symmetry imposed nodal behavior
such as in the case of the $d$-wave state. Although the
$T$-dependence of $T_{1}^{-1}$ in the vortex state is much like that
of the $s_{\pm}$-wave case, it is less influenced by the impurities
[see Fig. 5].

\vspace*{-.2cm}
\section {conclusion}
In conclusion, we have investigated the effect of the magnetic field
and nonmagnetic impurities on the DOS and the spin-lattice
relaxation rate $T_{1}^{-1}$ in the iron-based superconductors. It
is shown that the characteristic site-averaged DOS depends on the
gap functions in the presence of magnetic vortices and nonmagnetic
impurities. The magnetic vortices have a significant contribution to
the spin-lattice relaxation rate and lead to the relation
$T_{1}^{-1}\sim T^{3}$ followed by a nearly $T$-linear dependence at
low temperatures for all three symmetries ($s_{\pm}$-, $s$- and
$d$-wave) considered here, though in the clean uniform state a
$T^{3}$ dependence for the $d$-wave symmetry differentiates from the
others with a $T^{5}$ dependence. In the presence of nonmagnetic
impurities, this enhancement of $T_{1}^{-1}$ can only be seen for
the $s_{\pm}$- and $d$-wave symmetries, whereas it is almost
unaffected for the $s$-wave symmetry. Our results suggest that it is
impossible to distinguish the $s_{\pm}$- and $d$-wave symmetries on
the basis of the measurements of spin-lattice relaxation rates when
a magnetic field and/or impurity scatterings are present.

\section{acknowledgement}
\par This work was supported by the National Natural Science Foundation
of China (Grant No. 10904062 and No. 91021001), Hangzhou Normal
University (HSKQ0043, HNUEYT), and the Ministry of Science and
Technology of China (973 project Grants Nos. 2011CB922101,
2011CB605902).

\vspace*{.2cm}
\begin{figure}[htb]
\begin{center}
\vspace{-.2cm}
\includegraphics[width=240pt,height=210pt]{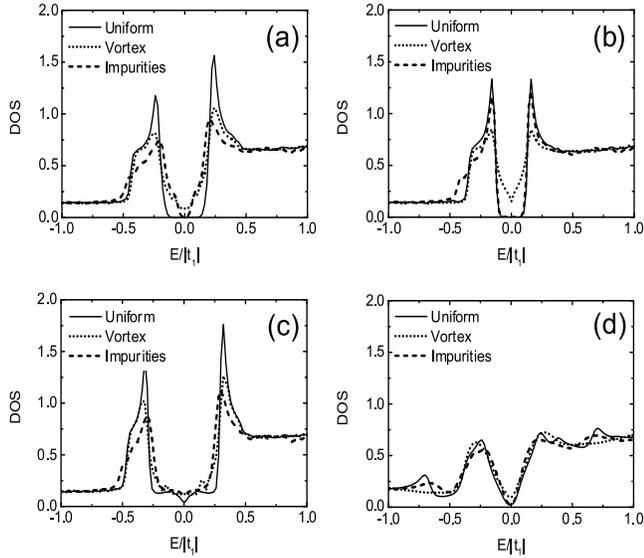}
\caption{Site-averaged DOS for the $s_{\pm}$-wave in (a), the
$s$-wave in (b), the $d$-wave in (c), and the $s_{h}$-wave in (d)
(see text). The DOS in the uniform SC state is plotted with solid
lines. The results in the magnetic vortex state and those in the
presence of nonmagnetic impurities are shown with the dotted and
dashed lines, respectively.}\label{fig1}
\end{center}
\end{figure}
\vspace*{-.0cm}
\begin{figure}[htb]
\begin{center}
\vspace{-.2cm}
\includegraphics[width=240pt,height=110pt]{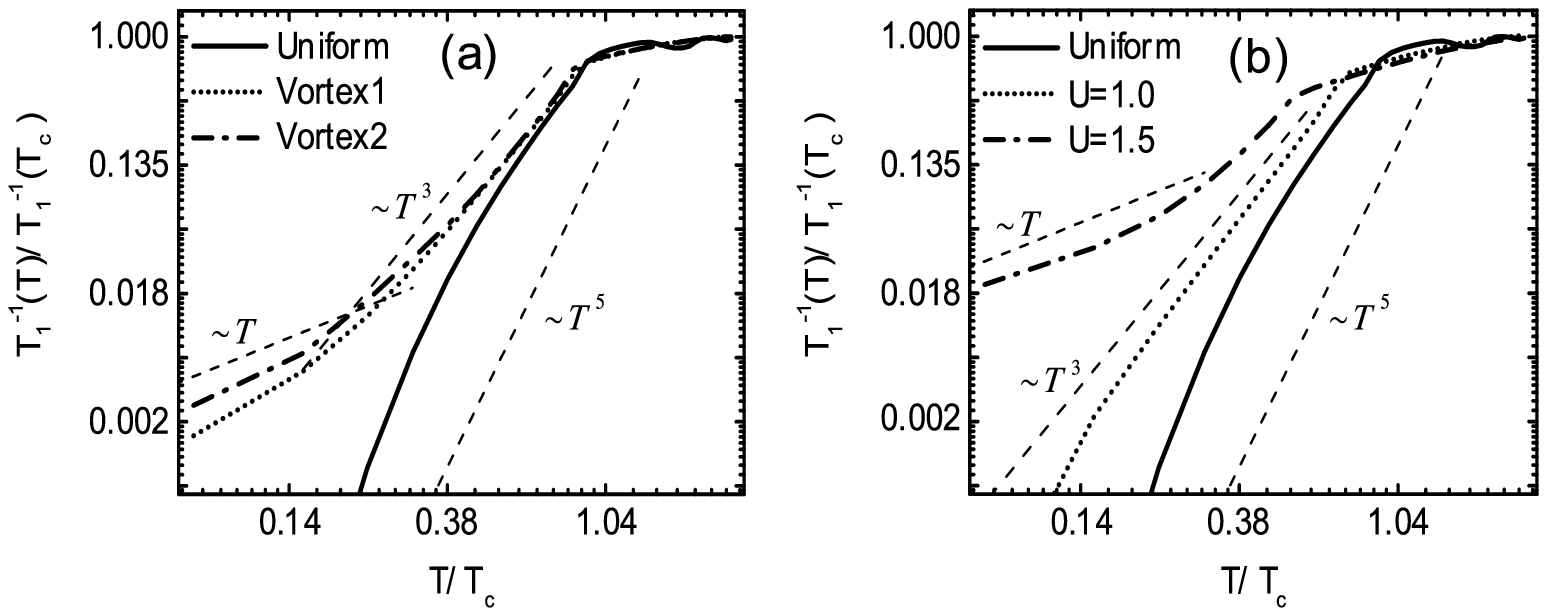}
\caption{$T$-dependence of $T_{1}^{-1}$ shown in the double
logarithmic chart for $s_{\pm}$-wave symmetry in the vortex state
(a), and in the presence of nonmagnetic impurities (b). The dotted
and dash-dotted curves denote different strength of the magnetic
field (see text) and the impurity scattering. The results for the
uniform SC state are also plotted with solid line in each
figure.}\label{fig2}
\end{center}
\end{figure}
\vspace*{-.0cm}
\begin{figure}[htb]
\begin{center}
\vspace{-.2cm}
\includegraphics[width=240pt,height=110pt]{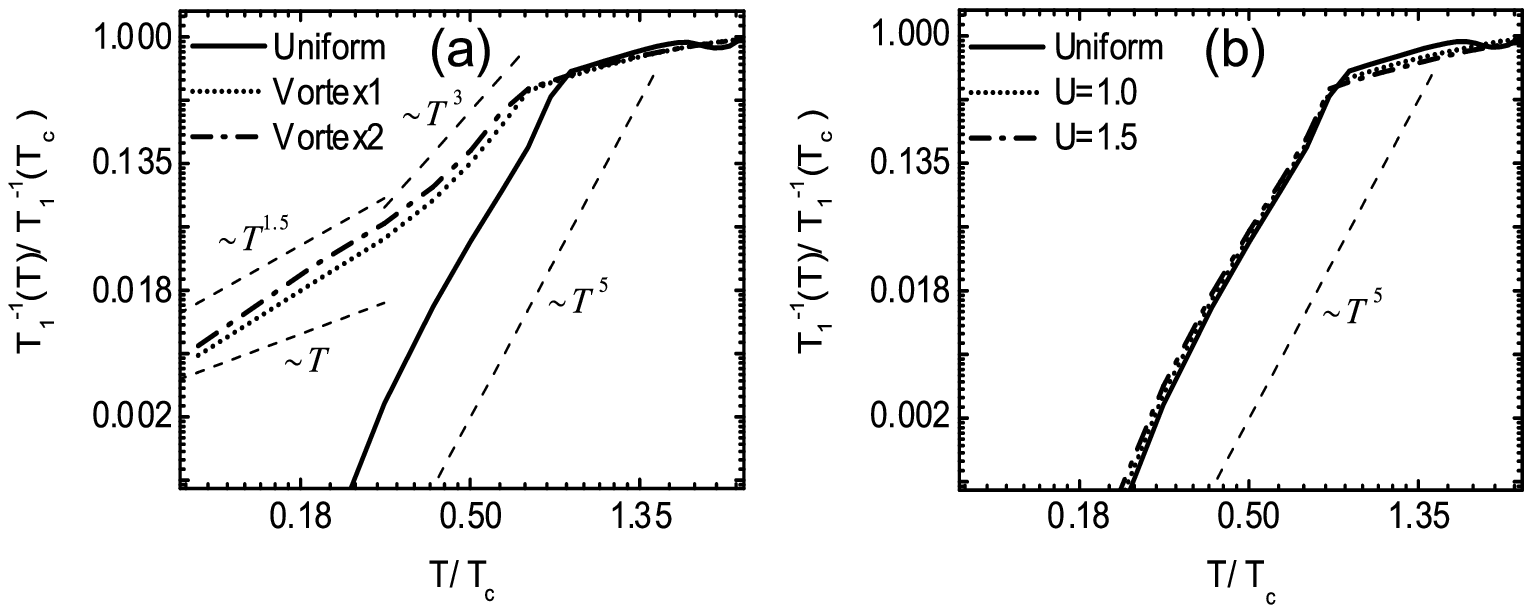}
\caption{$T$-dependence of $T_{1}^{-1}$ shown in the double
logarithmic chart for $s$-wave symmetry in the vortex state (a), and
in the presence of nonmagnetic impurities (b). The dotted and
dash-dotted curves denote different strength of the magnetic field
(see text) and the impurity scattering. The results for the uniform
SC state are also plotted with solid line in each
figure.}\label{fig3}
\end{center}
\end{figure}
\vspace*{-.0cm}
\begin{figure}[htb]
\begin{center}
\vspace{-.2cm}
\includegraphics[width=240pt,height=110pt]{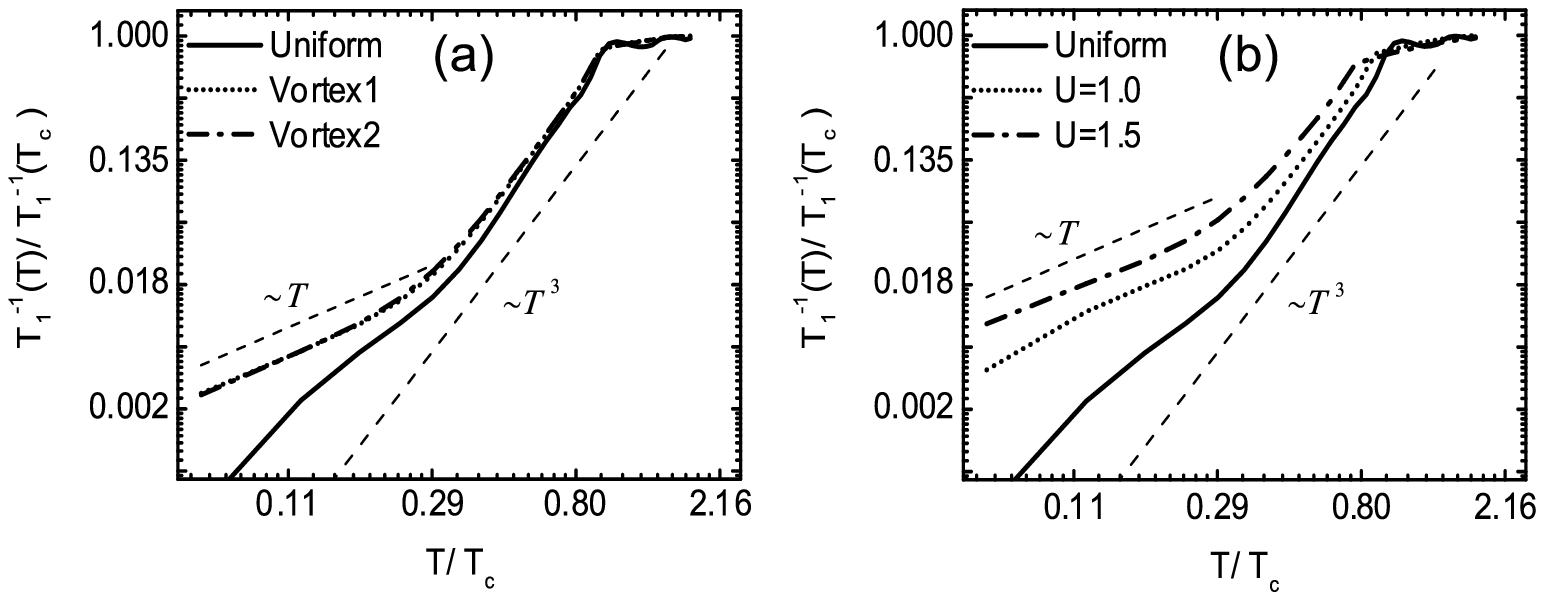}
\caption{$T$-dependence of $T_{1}^{-1}$ shown in the double
logarithmic chart for $d$-wave symmetry in the vortex state (a), and
in the presence of nonmagnetic impurities (b). The dotted and
dash-dotted curves denote different strength of the magnetic field
(see text) and the impurity scattering. The results for the uniform
SC state are also plotted with solid line in each
figure.}\label{fig4}
\end{center}
\end{figure}
\vspace*{-.0cm}
\begin{figure}[htb]
\begin{center}
\vspace{-.2cm}
\includegraphics[width=240pt,height=110pt]{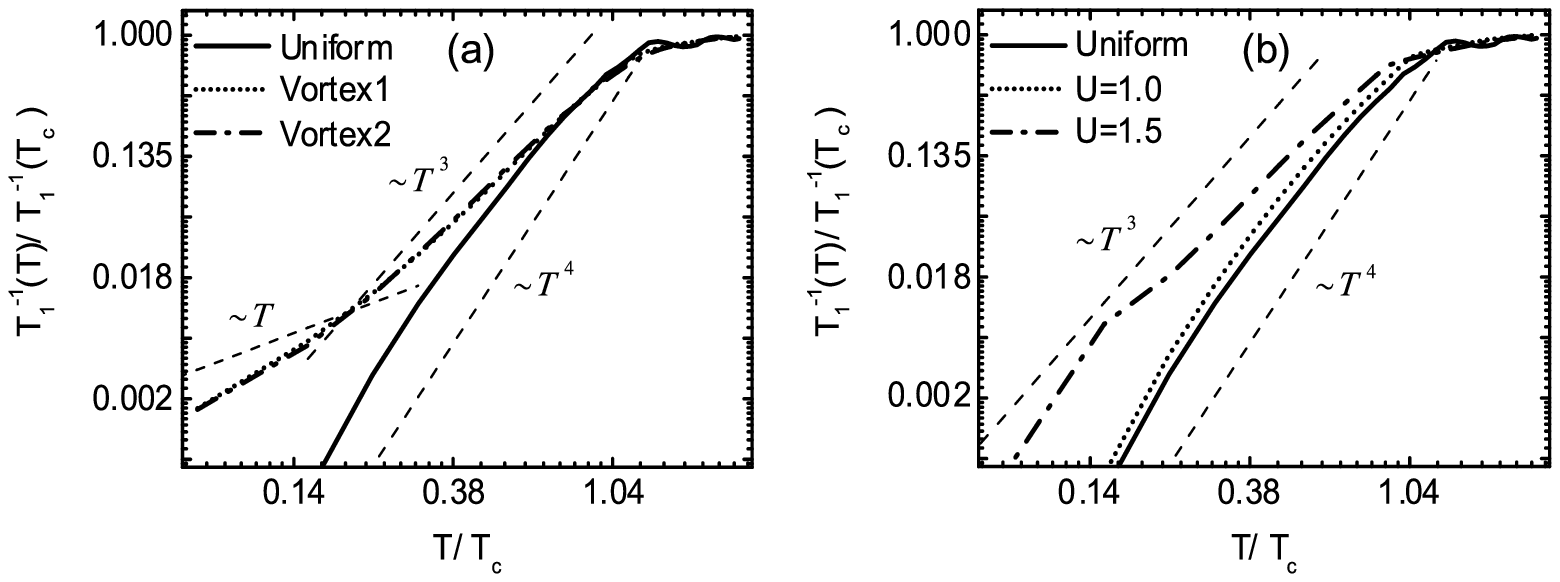}
\caption{$T$-dependence of $T_{1}^{-1}$ shown in the double
logarithmic chart for $s_{h}$-wave symmetry (see text) in the vortex
state (a), and in the presence of nonmagnetic impurities (b). The
dotted and dash-dotted curves denote different strength of the
magnetic field (see text) and the impurity scattering. The results
for the uniform SC state are also plotted with solid line in each
figure.}\label{fig5}
\end{center}
\end{figure}
\vspace*{-.0cm}

\end{document}